\documentclass[%
reprint,
superscriptaddress,
amsmath,amssymb,amsthm,mathrsfs,
aps,
prl,
]{revtex4-1}

\usepackage{graphicx}
\usepackage{dcolumn}
\usepackage{bm}
\usepackage[usenames]{color}
\usepackage[ 
margin=0.75in,
]{geometry}
\usepackage[normalem]{ulem}

\def\SNFA{Sr$_{1-x}$Na$_{x}$Fe$_2$As$_2$}
\def\xtwo{Sr$_{0.71}$Na$_{0.29}$Fe$_2$As$_2$}
\def\xthree{Sr$_{0.66}$Na$_{0.34}$Fe$_2$As$_2$}
\def\TN{$T_\mathrm{N}$}
\def\Tc{$T_c$}
\def\Tr{$T_r$}
\def\Ts{$T_s$}
\def\CF{$C_4$}
\def\CT{$C_2$}

\def\rmin{$r_{\mathrm{min}}$}

\def\rmid{$r_{\mathrm{mid}}$}

\begin{document}

\title{
Local orthorhombicity in the magnetic \CF\ phase of the hole-doped iron-arsenide superconductor \SNFA
}

\author{Benjamin A. Frandsen}
\affiliation{%
	Materials Sciences Division, Lawrence Berkeley National Laboratory, Berkeley, California 94720, USA.
}%
	\affiliation{ %
	Department of Physics, University of California, Berkeley, California 94720, USA.
} %

\author{Keith M. Taddei}
\affiliation{%
	Quantum Condensed Matter Division, Oak Ridge National Laboratory, Oak Ridge, Tennessee 37831, USA.
}%

\author{Ming Yi}
\affiliation{ %
	Department of Physics, University of California, Berkeley, California 94720, USA.
} %

\author{Alex Frano}
\affiliation{ %
	Department of Physics, University of California, Berkeley, California 94720, USA.
} %
\affiliation{
	Advanced Light Source, Lawrence Berkeley National Laboratory, Berkeley, California 94720, USA.
}

\author{Zurab Guguchia}
\affiliation{ %
	Department of Physics, Columbia University, New York, New York 10027, USA.
} %

\author{Rong Yu}
\affiliation{Department of Physics, Renmin University of China, Beijing 100872, China.}
\affiliation{Department of Physics and Astronomy, Shanghai Jiao Tong
	University, Shanghai 200240, China}

\author{Qimiao Si}
\affiliation{
	Department of Physics and Astronomy \& Rice Center for Quantum Materials, Rice University, Houston, Texas 77005, USA.
}

\author{Daniel E. Bugaris}
\affiliation{%
	Materials Science Division, Argonne National Laboratory, Argonne, Illinois 60439, USA.
}%

\author{Ryan Stadel}
\affiliation{ %
	Department of Physics, Northern Illinois University, DeKalb, Illinois 60439, USA.
} %
\affiliation{%
	Materials Science Division, Argonne National Laboratory, Argonne, Illinois 60439, USA.
}%

\author{Raymond Osborn}
\affiliation{%
	Materials Science Division, Argonne National Laboratory, Argonne, Illinois 60439, USA.
}%

\author{Stephan Rosenkranz}
\affiliation{%
	Materials Science Division, Argonne National Laboratory, Argonne, Illinois 60439, USA.
}%

\author{Omar Chmaissem}
\affiliation{ %
	Department of Physics, Northern Illinois University, DeKalb, Illinois 60439, USA.
} %
\affiliation{%
	Materials Science Division, Argonne National Laboratory, Argonne, Illinois 60439, USA.
}%

\author{Robert J. Birgeneau}
\affiliation{ %
	Department of Physics, University of California, Berkeley, California 94720, USA.
} %
\affiliation{%
	Materials Science Division, Lawrence Berkeley National Laboratory, Berkeley, California 94720, USA.
}%
\affiliation{ %
	Department of Materials Science and Engineering, University of California, Berkeley, California 94720, USA.
} %

\begin{abstract}
We report temperature-dependent pair distribution function measurements of \SNFA, an iron-based superconductor system that contains a magnetic phase with reentrant tetragonal symmetry, known as the magnetic \CF\ phase. Quantitative refinements indicate that the instantaneous local structure in the \CF\ phase is comprised of fluctuating orthorhombic regions with a length scale of $\sim$2~nm, despite the tetragonal symmetry of the average static structure. Additionally, local orthorhombic fluctuations exist on a similar length scale at temperatures well into the paramagnetic tetragonal phase. These results highlight the exceptionally large nematic susceptibility of iron-based superconductors and have significant implications for the magnetic \CF\ phase and the neighboring \CT\ and superconducting phases.
\end{abstract}

\maketitle


Much of the effort to understand high-temperature superconductivity focuses on the electronic phases occurring in close proximity to the superconducting (SC) state, since these orders and their fluctuations are expected to be closely related to the SC mechanism~\cite{pagli;np10,keime;n15}. Among the iron-based superconductors, nearly all parent compounds exhibit coupled structural and magnetic transitions from a paramagnetic tetragonal phase at high temperature to an antiferromagnetic (AF) orthorhombic phase at low temperature. The magnetic transition occurs at the same or slightly lower temperature than the structural transition ~\cite{dai;rmp15}. This orthorhombic phase has been classified as an electronic nematic state on the basis of its highly anisotropic electronic properties, which cannot be readily explained by the relatively small structural distortion alone~\cite{ferna;np14}. Optimal SC typically emerges when the AF orthorhombic state is suppressed by doping, chemical pressure, or hydrostatic pressure, suggesting that it is intimately related to the SC state~\cite{chen;nsr14}.


Recently, a magnetic phase with reentrant tetragonal symmetry was discovered in several hole-doped materials, including Ba$_{1-x}$Na$_{x}$Fe$_2$As$_2$~\cite{avci;nc14}, Ba$_{1-x}$K$_{x}$Fe$_2$As$_2$~\cite{bohme;nc15}, Sr$_{1-x}$Na$_{x}$Fe$_2$As$_2$~\cite{tadde;prb16}, and Ca$_{1-x}$Na$_{x}$Fe$_2$As$_2$~\cite{tadde;prb17}. For a small composition range, these systems undergo successive transitions from paramagnetic tetragonal to AF orthorhombic to AF tetragonal phases as the temperature is lowered. Due to its four-fold rotational invariance, this AF tetragonal phase is called the magnetic \CF\ phase, distinct from the usual magnetic \CT\ phase. The magnetic order in the \CF\ structure differs significantly from that in the \CT\ phase, exhibiting an out-of-plane spin reorientation~\cite{wasse;prb15,allre;prb15} and a coherent superposition of two orthogonal spin density waves~\cite{malle;epl15,allre;np16}, whereas the \CT\ phase selects just one of those wavevectors. The apparent universality of this phase in hole-doped iron pnictides has rekindled the discussion on the relationship between nematicity and SC in iron-based superconductors. Determining the microscopic nature of this surprising \CF-symmetric phase is one of the major challenges in the field of iron pnictides. Furthermore, the relationship among the \CF\ phase and its neighboring phases has bearing on the more general phenomenon of unexpected behaviors emerging out of competing orders, a recurring theme for unconventional superconductors and strongly correlated electron systems.

Up to this point, information about the atomic structure has been acquired by techniques such as standard x-ray/neutron powder diffraction and capacitive dilatometry that are sensitive to the \textit{average} crystallographic structure, but do not directly probe the \textit{local} atomic structure on short length scales. In contrast, pair distribution function (PDF) analysis of x-ray/neutron total scattering measurements can be used to investigate quantitatively the local atomic structure, which often differs from the average structure and can be very influential in determining the underlying microscopic physics~\cite{egami;b;utbp12}. PDF measurements have offered key insights into the properties of many families of strongly correlated electron systems, such as local polaron formation in the manganites~\cite{billi;prl96,bozin;prl07,shatn;prb16}, inhomogeneous electronic states in the pseudogap phase of the cuprates~\cite{bozin;prl00,bozin;prb15}, and fluctuating charge stripes in the nickelates~\cite{abeyk;prl13}.

A typical PDF experiment involves Fourier transforming the energy-integrated total scattering intensity (including Bragg and diffuse scattering) up to a large momentum transfer to obtain the corresponding structural information directly in real space. For PDF data acquired on neutron time-of-flight diffractometers, the inelastic scattering contributes structural information to the PDF up to an instrument-dependent effective maximum energy transfer~\cite{egami;b;utbp12}. A reasonable lower-bound estimate for this maximum is tens of meV. The PDF reveals the equal-time (i.e., instantaneous) local atomic correlations within the corresponding time window, or about 10$^{-13}$~s for 10~meV, in contrast to the static correlations probed by Bragg scattering. 
To date, relatively few PDF studies of iron pnictides~\cite{josep;jpcm11,malav;jpcm11,niedz;prb12,marti;jssc14,josep;jpcs15} and iron chalcogenides~\cite{louca;prb10,carr;prb14} have been reported, and in many cases without any availability of systematic temperature- or composition-dependent data. Nevertheless, these works have provided indications that the local structure  of these materials can differ from the average structure in important ways, suggesting that further study is worthwhile. The PDF measurements we report here have the benefit of a detailed temperature dependence spanning multiple structural phases for multiple compositions, providing unprecedented insights into the evolution of the local atomic structure in iron-based superconductors.

\begin{figure}
	\includegraphics[width=60mm]{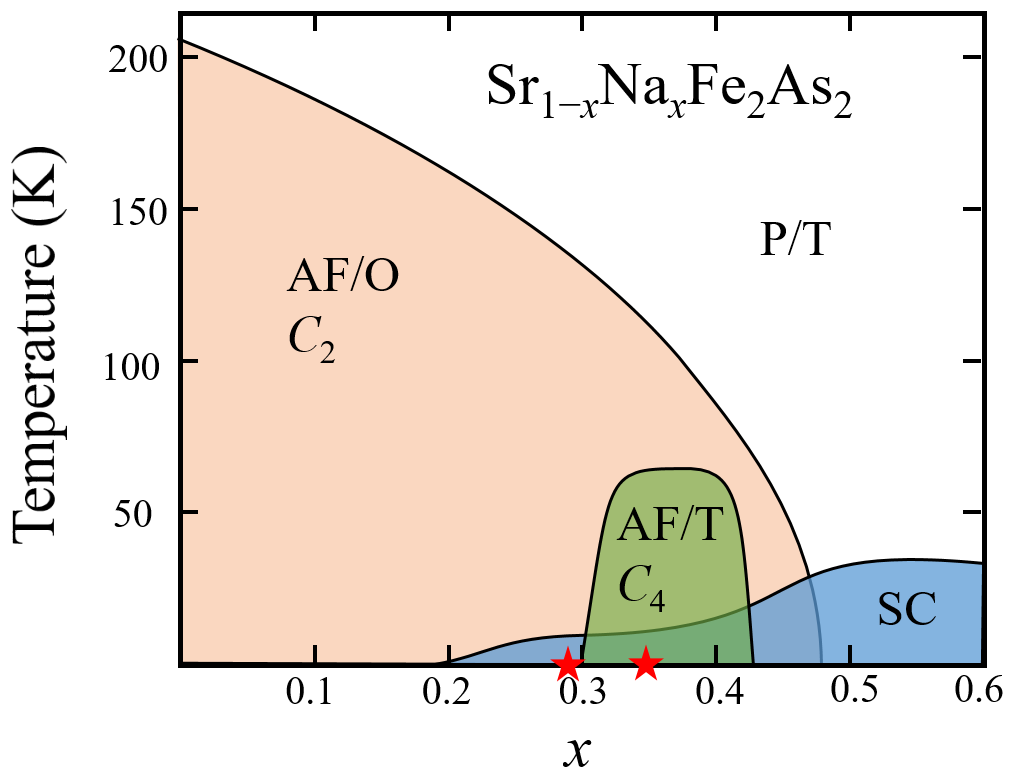}
	\caption{\label{fig:PD} (Color online) Temperature versus composition phase diagram of \SNFA, showing the paramagnetic tetragonal (P/T), antiferromagnetic orthorhombic (AF/O), antiferromagnetic tetragonal (AF/T), and superconducting (SC) phases. Adapted from Ref.~\onlinecite{tadde;prb16}.}
	
\end{figure}

To determine the instantaneous local structure in and around the magnetic \CF\ phase, we performed PDF analysis of temperature-dependent neutron scattering measurements of two compositions of \SNFA\ (see Fig.~\ref{fig:PD} for the phase diagram).
The \CF\ phase for \SNFA\ has the largest known compositional extent and highest known transition temperature \Tr, making it ideal for this study. We present two primary findings from our investigation. First, the instantaneous local structure in the \CF\ phase consists of fluctuating orthorhombic regions with a length scale of approximately 2~nm, with no reduction of orthorhombicity across \Tr\ on this length scale. Second, for compositions inside and outside the \CF\ range, the instantaneous atomic structure develops significant orthorhombic distortions at temperatures well above the long-range \CT\ structural phase transition at \Ts. Both of these findings are manifestations of the large nematic susceptibility of these materials, resulting in robust orthorhombicity in the instantaneous local structure even when the average structure is tetragonal. These results resolve the mystery of the atomic structure in the magnetic \CF\ phase and have significant implications for the complex relationship among the SC, \CT, and \CF\ phases.

We measured powder specimens of \SNFA\ with $x = 0.29$ and 0.34 (shown as red stars on the horizontal axis in Fig.~\ref{fig:PD}), both of which were characterized by standard neutron diffraction in Ref.~\onlinecite{tadde;prb16}. The $x = 0.29$ sample is located in the phase diagram just to the left of the \CF\ dome, transitioning to the \CT\ phase at 140~K and remaining orthorhombic at all lower temperatures. The $x = 0.34$ sample is located near the center of the \CF\ region, with the initial \CT\ transition occurring at 100~K and the magnetic \CF\ phase forming below 70~K. All structural and magnetic transitions occur simultaneously and are first-order, and both samples become superconducting below \Tc~$\approx$~10~K~\cite{tadde;prb16}. Temperature-dependent neutron total scattering experiments were performed on the NOMAD beamline~\cite{neufe;nimb12} at the Spallation Neutron Source (SNS) of Oak Ridge National Laboratory (ORNL). The total scattering data were reduced and transformed with $Q_{\mathrm{max}}$ = 36~\AA$^{-1}$ using the automatic data reduction scripts at the NOMAD beamline, and the PDF modeling was carried out using the PDFgui program~\cite{farro;jpcm07}. Estimates of the standard uncertainty of the refined parameters of the least-squares fits were obtained according to the protocol in Appendix B of Ref.~\onlinecite{yang;jac14}.

\begin{figure}
	\includegraphics[width=80mm]{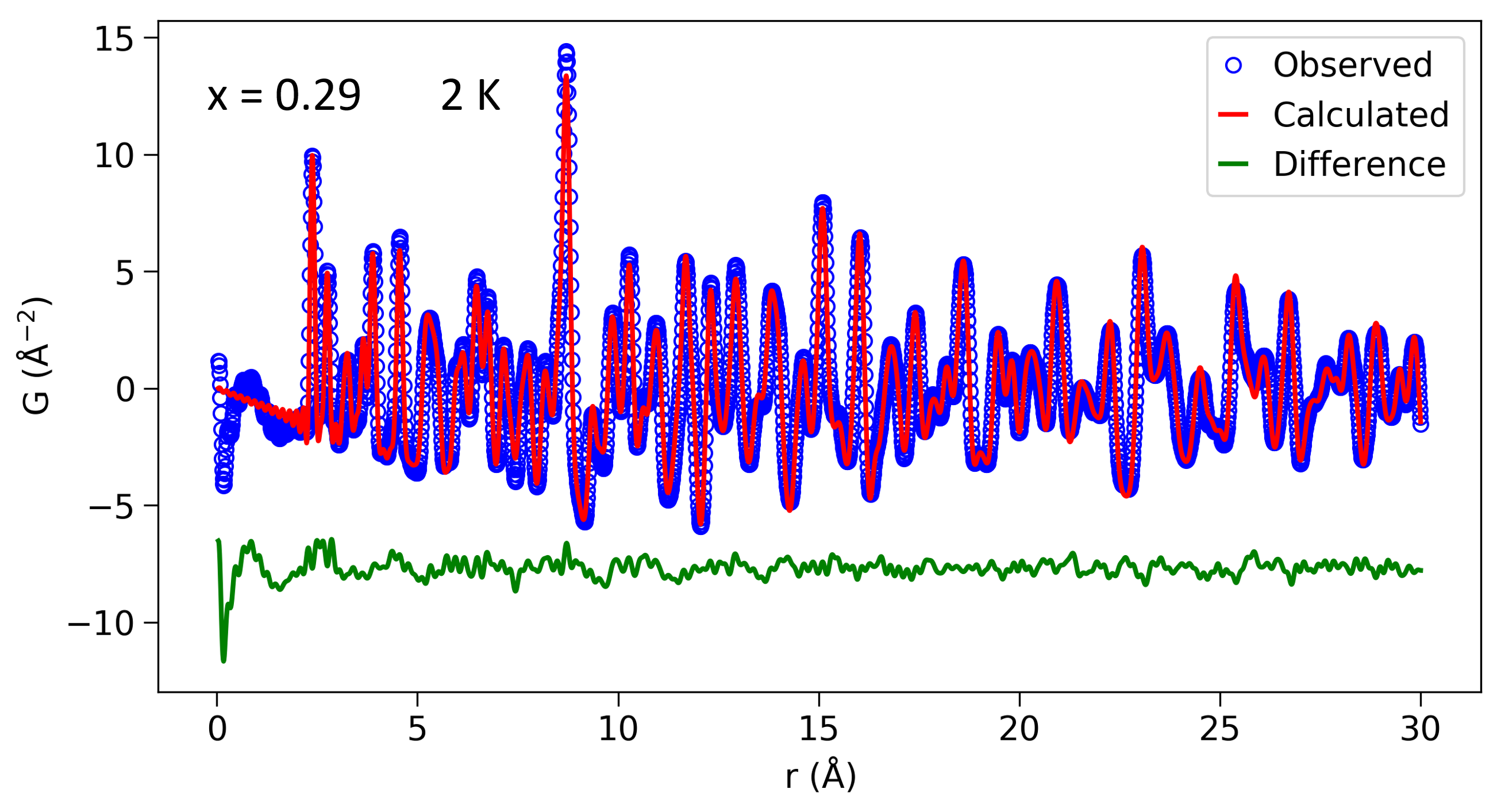}
	\caption{\label{fig:fit} (Color online) Typical PDF refinement of \xtwo\ at 2~K using the orthorhombic structural model. The blue circles represent the data, the red curve the best fit, and the green curve the fit residual, offset vertically for clarity.}
	
\end{figure}

The result of a typical PDF refinement for $x = 0.29$ at 2~K is displayed in Fig.~\ref{fig:fit}. 
Aside from the obvious distortions in the data at low $r$ ($\lesssim 3$~\AA) that commonly appear as artifacts from the Fourier transform, the data are quite well described by the PDF calculated from the refined orthorhombic structural model, with goodness-of-fit measures similar in quality to those for a standard Si sample measured under identical conditions. This confirms that the data are of high enough quality for quantitative structural refinement. Since both samples are magnetically ordered at low temperature, the magnetic pair distribution function (mPDF) is present in the data along with the atomic PDF~\cite{frand;aca14,frand;aca15}, but on a scale that is expected to be more than two orders of magnitude smaller than the atomic PDF due to the small ordered moment in these materials. Therefore, the mPDF can be safely ignored in our analysis.

\begin{figure}
	\includegraphics[width=80mm]{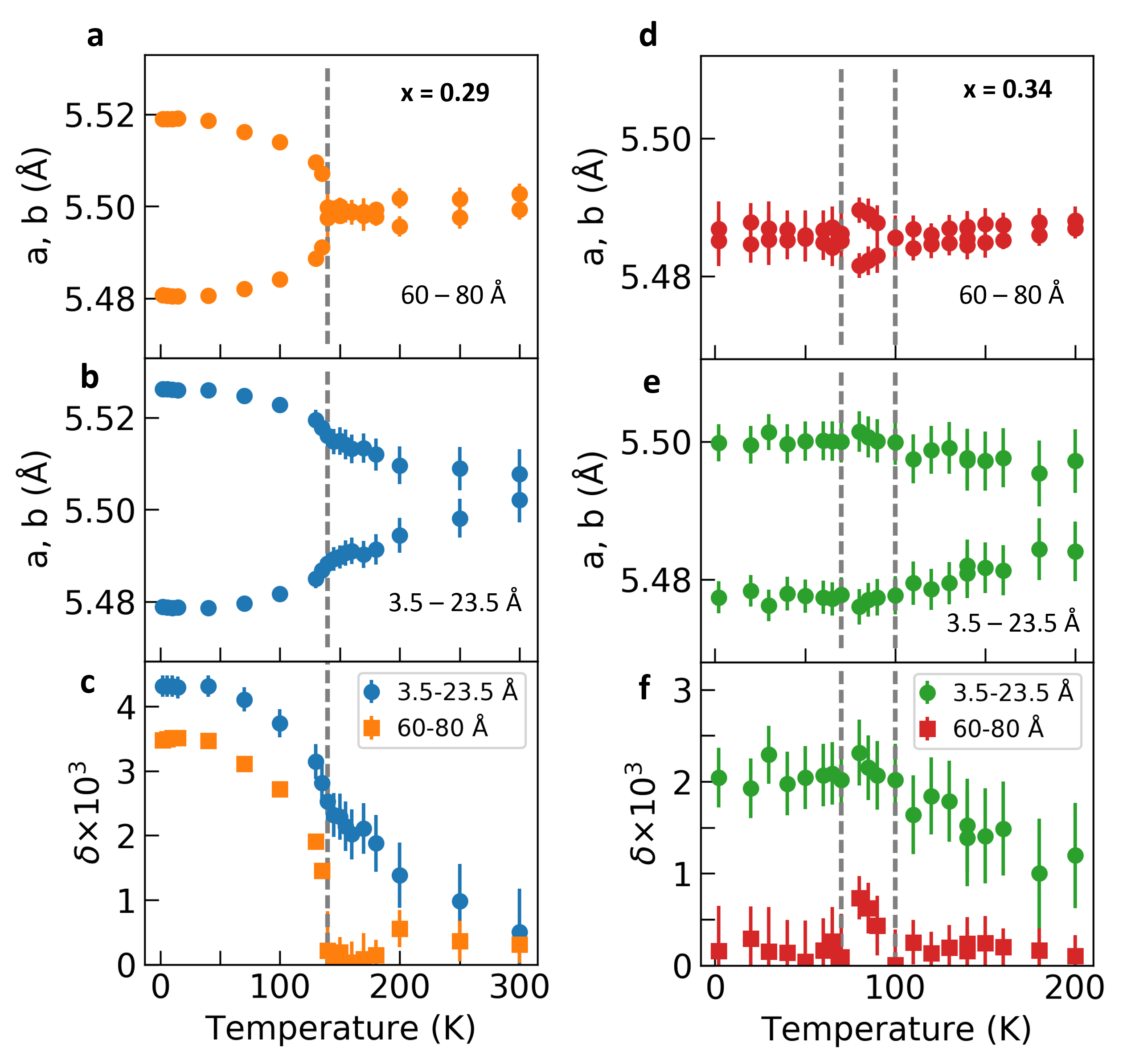}
	\caption{\label{fig:tempScan} (Color online) Refined \textit{a} and \textit{b} local lattice parameters as a function of temperature for different fitting ranges. Panels a and b show the results for \xtwo\ with fitting ranges of $60 \le r \le 80$~\AA and $3.5 \le r \le 23.5$~\AA, respectively. The vertical dashed line shows \Ts. Panel c displays the corresponding orthorhombicity for each fitting range. Panels d-f display the equivalent results for \xthree. The higher-temperature vertical dashed line indicates \Ts, the lower-temperature line \Tr.}
	
\end{figure}

To conduct a systematic investigation of the temperature-dependent structure on various length scales, we performed refinements with the orthorhombic model at all temperatures over two different fitting ranges, $3.5 \le r \le 23.5$~\AA\ and $60 \le r \le 80$~\AA. Fig.~\ref{fig:tempScan} displays the temperature dependence of the refined \textit{a} and \textit{b} lattice parameters extracted from these fits for both compositions. Although true crystallographic lattice parameters must be obtained from the positions of Bragg peaks, reflecting the average static structure, for convenience we will still refer to the refined values of $a$ and $b$ from the PDF fits as local lattice parameters.
The results for $x$ = 0.29 using the long fitting range, shown in panel (a), reveal a sharp transition at \Ts~=~140~K (indicated by the vertical dashed line) with a rapid orthorhombic splitting of \textit{a} and \textit{b} at lower temperatures and their convergence to values within the parameter uncertainties at higher temperatures, as expected for a tetragonal-to-orthorhombic transition. The splitting of \textit{a} and \textit{b} at low temperature is 0.039~\AA, similar to the 0.041~\AA\ splitting determined by earlier diffraction analysis~\cite{tadde;prb16}. Therefore, on a length scale of 60---80~\AA, the structure is fully consistent with the expected crystallographic structure.

We contrast this with the instantaneous \textit{local} structure determined from refinements over the short fitting range, shown in panel (b). At low temperature, the orthorhombic splitting of \textit{a} and \textit{b} is 0.046~\AA, somewhat greater than that for the long-range fits. More importantly, this orthorhombic splitting remains quite significant even as the temperature is raised through \Ts, with the local lattice parameters slowly converging with increasing temperature until they are just within the refinement uncertainty at 300~K. Panel (c) displays the corresponding orthorhombicity parameter $\delta = (a - b)/(a + b)$ for the two fitting ranges, with the high-temperature tail and overall enhancement clearly visible for the short fitting range. On a length scale of at least $\sim 20$~\AA, then, the instantaneous local structure of \xtwo\ is strikingly different than the average structure, with an enlarged orthorhombic distortion that persists well into the high-temperature tetragonal phase. 

Panels (d-f) of Fig.~\ref{fig:tempScan} show the corresponding results for the $x=0.34$ sample. Once again, the refinements over the long fitting range [panel (d)] are consistent with the expected behavior for the average structure: \textit{a} and \textit{b} are equal to within the refinement uncertainties at all temperatures except between  \Tr\ and \Ts\ (displayed by vertical dashed lines), where a small but unambiguous orthorhombic splitting is evident. The magnitude of this splitting (0.009~\AA) is in reasonable agreement with the earlier result from standard diffraction (0.012~\AA). On the other hand, analysis of the short fitting range in panel (e) reveals a nonzero orthorhombic splitting of \textit{a} and \textit{b} even at 200~K, which increases gradually as the temperature is lowered and remains nearly constant below \Ts. In stark contrast to the average structure, there is no recovery of tetragonal symmetry below \Tr\ for this fitting range: the instantaneous local structure remains unambiguously orthorhombic in the magnetic \CF\ phase. However, the instantaneous local orthorhombicity displayed in panel (f) remains largely flat below \Ts, in contrast to the significant increase below \Ts\ for the $x$ = 0.29 sample. Therefore, the impact of the \CF\ phase on the instantaneous local structure appears to be at most a suppression of any further increase of the local orthorhombic distortion. The tetragonal model produces fits of consistently lower quality, verifying the orthorhombic nature of the instantaneous local structure in the \CF\ phase. The Supplementary Information contains further discussion about the two models. Inspection of the $(\frac{1}{2},\frac{1}{2},1)$ magnetic Bragg peak (using the conventional tetragonal setting) directly in the scattering data reveals the expected temperature dependence below \Ts\ and \Tr\ for both compositions (see Supplementary Information). 

Considering the remarkable structural differences between the short and long fitting ranges, it would be illuminating to determine the length scale over which the orthorhombically enhanced instantaneous local structure crosses over to the average structure. To this end, we performed sliding-range fits at a few representative temperatures for each compound, using the fitting ranges $r_{\mathrm{min}} \le r \le r_{\mathrm{min}}+20$~\AA, with \rmin\ varying in integer steps from 2 to 65~\AA. For each refinement, the \textit{a} and \textit{b} lattice parameters were extracted and used to compute the orthorhombicity. 

\begin{figure}
	\includegraphics[width=80mm]{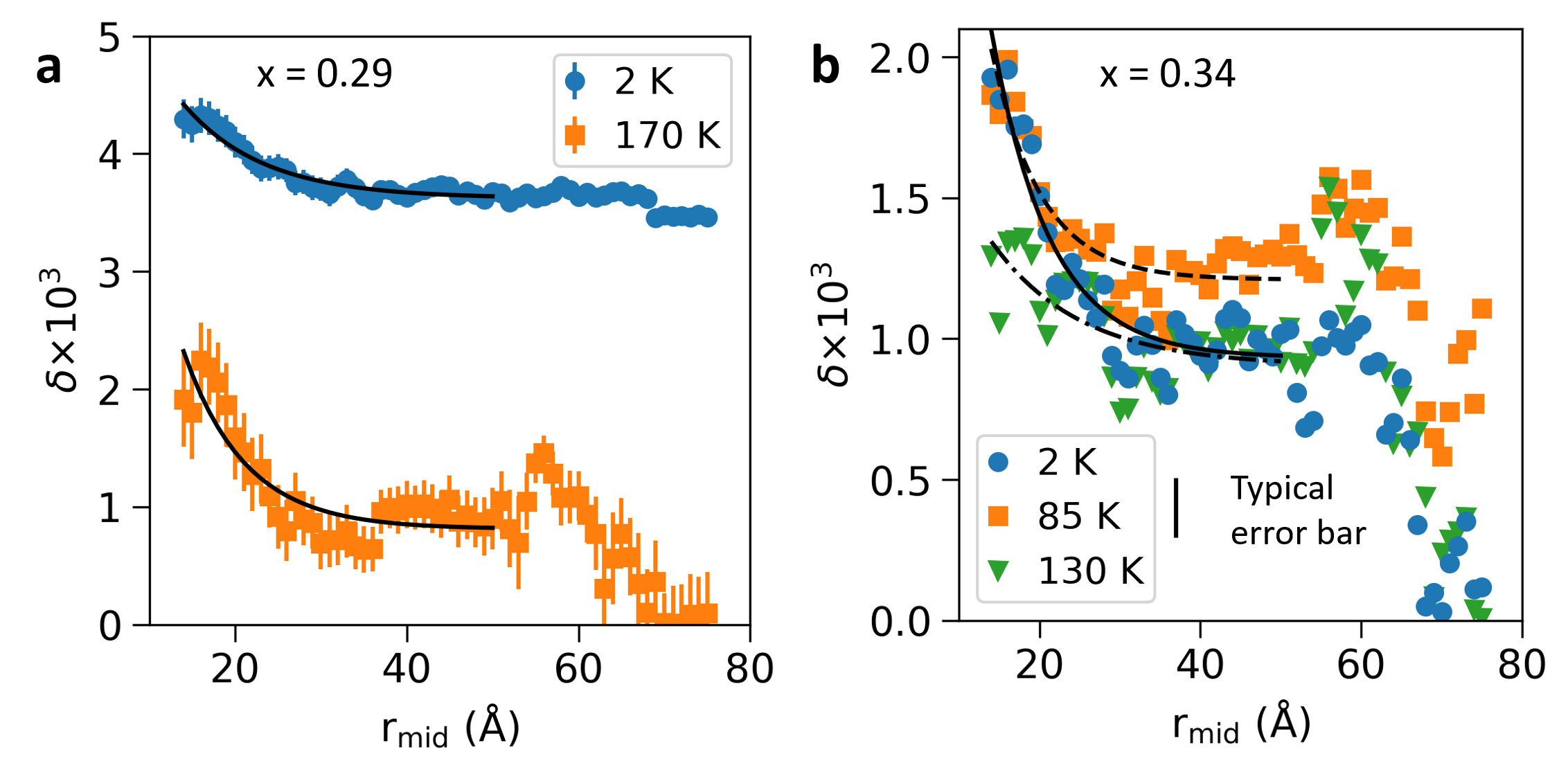}
	\caption{\label{fig:rScan} (Color online) Orthorhombicity as a function of fitting range for \xtwo\ (a) and \xthree\ (b) at various temperatures. The horizontal axis displays the midpoint of the sliding 20-\AA\ fitting range corresponding to each data point. Black curves show fits to the data using an exponential function plus a constant.}
	
\end{figure}

Fig.~\ref{fig:rScan}(a) displays the orthorhombicity as a function of the midpoint of the fitting range \rmid\ for $x$ = 0.29 at 2~K (deep inside the orthorhombic phase) and 170~K (in the paramagnetic phase).
At 2~K, the orthorhombicity is largest for the shortest fitting ranges and decreases gradually to plateau beyond approximately \rmid~=~30~\AA\ at a value of 0.0037, in good agreement with the earlier neutron diffraction analysis. Fitting an exponential function plus a constant [black curve in Fig.\ref{fig:rScan}(a)] yields an exponential decay length of 9.2 $\pm$ 1.2~\AA, corresponding to \rmid~$\approx$~20~\AA. At 170~K, the orthorhombicity is likewise maximal for short fitting ranges and decreases with a similar decay length (7.2 $\pm$ 1.5~\AA). As the fitting range increases, the orthorhombicity behaves somewhat non-monotonically, with a small peak around \rmid~=~56~\AA, followed by a final decrease to zero around \rmid~=~70~\AA. The instrumental resolution effects of the NOMAD beamline on the PDF are known to be difficult to model accurately for $r \gtrsim 50$~\AA, so the non-monotonic orthorhombicity at large $r$ is likely an artifact of these instrumental effects. This is discussed further in the Supplementary Information. Nevertheless, there is a clear trend at both temperatures for the orthorhombicity to be greatest at low $r$ and decrease gradually until it is significantly diminished for $r_{\mathrm{mid}}\gtrsim 20$~\AA.

Fig.~\ref{fig:rScan}(b) displays the equivalent data for the $x$ = 0.34 sample at 2~K (\CF\ phase), 85~K (\CT\ phase), and 130~K (paramagnetic phase). For clarity, the typical magnitude of the uncertainty is indicated separately rather than on each datum point. The results display significantly more scatter than those for $x$ = 0.29, but there are clear similarities: for all three temperatures, the orthorhombicity is greatest at low \textit{r}, relaxes over approximately 20~\AA, displays some non-monotonic behavior around \rmid~=~50 - 60~\AA\ attributable to instrumental resolution effects, and finally decreases to the average structure value (0.001 for 85~K, 0 for 2~K and 130~K) for \rmid\ around 70~\AA. At low \textit{r}, the results for 2~K are essentially identical to those for 85~K. However, as the fitting range increases beyond \rmid~=~40~\AA, the orthorhombicity at 2~K crosses over to match that at 130~K, providing direct evidence of the short-range nature of the instantaneous orthorhombic structure in the \CF\ phase. Fits yield decay lengths of 7.2 $\pm$ 1.0~\AA, 6.2 $\pm$ 1.2~\AA, and 11.0 $\pm$ 5.4~\AA\ for 2~K, 85~K, and 130~K, respectively, shown as solid, dashed, and dashed-dotted black curves in Fig.~\ref{fig:rScan}(b).

The main conclusion to be drawn from the sliding-range fits is that regardless of temperature and composition, the instantaneous local orthorhombicity decreases smoothly until it is largely suppressed for fits beyond the 10 - 30~\AA\ fitting range. Taking the midpoint of this range, the characteristic length scale of the instantaneous local structure is $\sim$20~\AA, which encompasses multiple unit cells in each direction. There may be a second length scale of $\sim$70~\AA\ corresponding to the fitting range where the PDF refinements agree with the standard neutron diffraction refinements, but this is less certain, given the instrumental resolution effects on the high-$r$ PDF data.


We also performed x-ray PDF measurements at Brookhaven National Laboratory (BNL). The data are consistent with short-range, instantaneous orthorhombicity developing well above \TN\ and persisting through the \CF\ phase at lower temperature. Due to the more limited range of momentum transfer available at the x-ray beamline compared to the neutron beamline, the x-ray PDF data do not yield any new information.

These PDF measurements provide important new insights into the atomic structure of \SNFA, revealing that the instantaneous local structure on a length scale of $\sim$20~\AA\ differs significantly from the average structure in all phases. In the \CT\ phase, the instantaneous local structure exhibits enhanced orthorhombicity relative to the average structure. In the tetragonal phases, the instantaneous local structure is not actually tetragonal at all, but is instead comprised of orthorhombic regions with a characteristic length scale of $\sim$20~\AA. The PDF data alone cannot determine whether these local orthorhombic regions are static or dynamic. However, earlier work using M\"ossbauer spectroscopy, a highly local probe in real space but with a relatively slow characteristic measurement time scale of $\sim$10$^{-7}$~s, detected only tetragonal symmetry in the \CF\ and paramagnetic tetragonal phases~\cite{allre;np16}. Therefore, we conclude that the orthorhombic regions in the \CF\ phase must be fluctuating on a time scale between 10$^{-7}$~s and 10$^{-13}$~s. We expect dynamic fluctuations in the high-temperature paramagnetic phase, as well. This would consistent with signatures of fluctuating orthorhombicity in the paramagneic phase in SrFe$_2$ As$_2$ detected by inelastic x-ray techniques~\cite{kobay;prb11,murai;prb16}. The novelty of the PDF measurements lies in their unique ability to probe the length scale of these fluctuations.

Our finding that the short-range, fluctuating orthorhombicity in the magnetic \CF\ phase is as pronounced as that in the \CT\ phase is an unexpected result. It suggests that the underlying nematic degrees of freedom are prevalent across these phases. Such microscopic nematic degrees of freedom should unify the descriptions of both phases and, indeed, a Ginzburg-Landau theory in this spirit is able to account for the observed dynamic local orthorhombicity in the magnetic \CF\ phase through direct calculation of the nematic susceptibility, which is found to be enhanced in the \CF\ phase~\cite{yu;arxiv17}. An alternative approach starting from an interacting electronic Hamiltonian may be found in Ref.~\onlinecite{wang;prb15}. We find it significant that the \CT\ phase already exists in the parent system, but the magnetic \CF\ phase only arises near the optimal doping for superconductivity. These results therefore provide evidence that nematic correlations are strong near optimal doping and should thus be closely related to superconductivity.

Our findings may have more general implications for iron-based superconductivity, as well. The strong tendency of the instantaneous local structure to undergo orthorhombic distortion above the \CT\ transition is consistent with the large nematic susceptibility known to exist in the paramagnetic phase~\cite{chu;s12,kuo;s16}. It seems likely that the dynamic local orthorhombicity observed at high temperature in \SNFA\ would also be found in other iron-pnictide and -chalcogenide materials. Indeed, this behavior is reminiscent of signatures of high-temperature nematicity observed in other systems, such as NaFeAs probed by scanning tunneling microscopy~\cite{rosen;np14} and BaFe$_2$(As,P)$_2$ via magnetic torque measurements~\cite{kasah;n12}, although the time scales probed by these techniques vary widely. Further PDF studies of these and other families of iron-based superconductors promise to yield rich information about the nature of the nematic phase. 
 
\begin{acknowledgments}
We thank Matt Tucker, Kate Page, and Marshall McDonnell for assistance at ORNL, and Eric Dooryhee, Milinda Abeykoon, Simon Billinge, and Timothy Liu for assistance at BNL. This work was performed primarily at Lawrence Berkeley National Laboratory and the University of California, Berkeley with support from the Office of Science, Office of Basic Energy Sciences (BES), Materials Sciences and Engineering Division, of the US Department of Energy (DOE) under Contract No. DE-AC02-05-CH11231 within the Quantum Materials Program (KC2202) and BES, US DOE, Grant No. DE-AC03-76SF008. Work at the Materials Science Division at Argonne National Laboratory was supported by the US DOE, Office of Science, Materials Sciences and Engineering Division. Work at Rice was supported by the NSF Grant No.\ DMR-1611392, the Robert A.\ Welch Foundation Grant No.\ C-1411 and a QuantEmX grant from ICAM and the Gordon and Betty Moore Foundation through Grant No. GBMF5305. Z.G. acknowledges the financial support by the Swiss National Science Foundation (SNFfellowship P2ZHP2- 161980). Q.S. acknowledges the hospitality of University of California, Berkeley. Work at Renmin University was supported in part by the National Science Foundation of China Grant Nos. 11374361 and 11674392 and Ministry of Science and Technology of China, National Program on Key Research Project Grant number 2016YFA0300504 (R.Y.). Use of the SNS, ORNL, was sponsored by the Scientific User Facilities Division, BES, US DOE. Use of the National Synchrotron Light Source II at Brookhaven National Laboratory, was supported by DOE-BES under Contract No. DE-SC0012704. 
\end{acknowledgments}

\end{document}